\documentclass[aps,physrev,twocolumn,groupedaddress, amsmath]{revtex4-2}

\usepackage{graphicx}
\usepackage{dcolumn}
\usepackage{bm}
\usepackage[colorlinks=true, linkcolor=blue, citecolor=blue, urlcolor=blue]{hyperref}

\begin{document}
\title{Constitutive flow law for hydrogel granular rafts near the brittle-ductile transition}

\author{Yuto Sasaki}
\email[]{sasaki.geoscience@gmail.com}
\author{Hiroaki Katsuragi}
\affiliation{Department of Earth and Space Science, The University of Osaka}

\begin{abstract}
Spatially varying flow laws have been identified in dry granular flow, yet their applicability to unjammed suspensions remains unclear.
This study demonstrates that the quasistatic suspension flow combines dry granular rheology with nonlocal effects in the shear band and damped viscous flow in the outer creep region.
Through rotary shear experiments on a hydrogel granular raft, we observe that the flow decays from the interface in the quasistatic regime, where the particles remain mobile even below the yield stress.
These findings suggest the universal flow law across the transition between jammed/brittle granular behavior and unjammed/ductile viscous flow.
\end{abstract}

\maketitle
\section{Introduction}\label{sec:intro}

Suspensions are a common type of multiphase system in which granular materials are dispersed in fluid media.
They are ubiquitous, as found in mudflows, pumice rafts, floating plastic debris, and powder manufacturing processes.
They are expected to exhibit flow characteristics that combine those of viscous fluids and dry particles.
Following the flow law of viscous fluids, i.e., the relationship between shear stress $\tau$ and strain rate $\dot{\gamma}$, the flow of apparently rigid dry granular systems has been found \cite{terada_experimental_1929} and extensively studied (e.g., \cite{gdr_midi_dense_2004}). 
The enduring-contact flow behavior of granular systems follows a power-law rate dependence with a yield strength $\tau_\mathrm{y}$ at sufficiently low strain rates, as $\tau= \tau_\mathrm{y}+k\dot{\gamma}^n$ with a flow consistency index $k$ and an exponent $n$ \cite{bird_rheology_1983}.
This flow law is successfully scaled using the friction $\mu=\tau/P$ and the inertial number $I=\dot{\gamma} d/\sqrt{P/\rho_\mathrm{p}}$, where $P$ is pressure, $d$ is particle diameter, and $\rho_\mathrm{p}$ is particle density, regardless of system details \cite{da_cruz_rheophysics_2005,jop_constitutive_2006}.

However, one striking difference from viscous fluids is that the flow law of local microscopic elements does not necessarily represent that of the bulk system \cite{goyon_spatial_2008}. This discrepancy is referred to as nonlocality or size effects in flow behavior (e.g., \cite{bouzid_non-local_2015}).
Such nonlocality is observed particularly in the quasistatic regime, defined in this study as the regime where shear is sufficiently slow compared to the inertial or viscous rearrangement of individual particles.
It has been pointed out that external disturbances, such as those from shear flow boundaries, play an important role.
Many models have been suggested, principally focusing steady diffusional equation of shearing activity \cite{bouzid_non-local_2015,gaume_microscopic_2020}.

Despite many efforts to describe the granular nonlocality,
its physical origin and the limitations of its applicability, as well as the explicit form of the flow law, remain controversial.
Most studies have investigated this unique property using dry granular systems,
mainly through numerical simulation \cite{koval_annular_2009, kamrin_nonlocal_2012,zhang_microscopic_2017,kamrin_non-locality_2019,kim_power-law_2020,gaume_microscopic_2020}.
However, the nonlocality in a suspension remains unexplored, which is difficult to investigate through numerical simulation and was found only by a few seminal experimental works \cite{huang_flow_2005}.
The transitional connection between fluids' local flow and granular nonlocal response is crucial for understanding the brittle-ductile transition mechanism of dense suspensions, the size effects of powder manufacturing, and the seismogenic fault yielding.

To investigate the nonlocal effects,
soft particles, e.g., solid hydrogels (this study), liquid emulsions \cite{goyon_spatial_2008}, and gas bubbles \cite{katgert_couette_2010}, are useful.
They enhance local interparticle sliding and global (visco)elastic relaxation, both of which are considered to induce the nonlocality \cite{goyon_spatial_2008, bouzid_non-local_2015}.
In addition to them, suspensions have inherent viscous drag by fluid as well,
suggesting that the nonlocality could significantly influence shear properties and shear-induced structures even below the jamming transition point \cite{liu_jamming_1998,behringer_physics_2019} in two dimensions (2D) \cite{lalieu_rheology_2023} and 3D \cite{huang_flow_2005}. 
To elucidate the unclear relationship between the microscopic local motion to the bulk flow law in suspensions,
a shear experiment of a monolayer granular raft floating on liquid surface is convenient to directly capture the particle motion and apply shear without basal friction \cite{lalieu_rheology_2023}.
Additionally, in a Couette cell geometry, the shear stress is uniquely determined by the radial distance, and the packing fraction can be easily varied within a constant volume, both of which control the flow behavior.
Therefore, the granular raft system sheared in a Couette cell is effective for investigating the flow law and the nonlocality mechanism working in suspensions.

In this study, we investigate the local flow laws of both wall and internal particles at various packing fractions and system sizes in the quasistatic regime, where the bulk system exhibits a yield strength.
Shear experiments were conducted on a quasi-2D granular raft in various Couette cells, measuring the torque and all particle motions.
We have found that the local particles flow even in the quasistatic regime.
Particle velocity exhibits a central shear band with exponentially decaying flow and a creep zone outside with slower decay.
When a diffusional equation for the inertial number $I$ is applied to the suspension,
the characteristic diffusion length leads to a universal flow law within the shear band.

\section{Methods}\label{sec:meth}
\begin{figure}
\includegraphics[width=87mm]{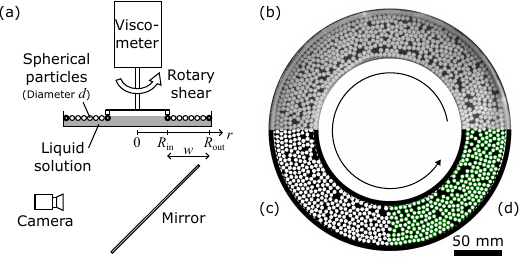}%
\caption{Experimental setup and image analysis. (a) Schematic of the experiment. The thick particles are fixed on the walls except for the widest-channel runs. (b--d) Representative particle image from Run \#99 showing the same field of view. The image is divided into three panels: raw (b), binary (c), and particle-detection images (d). Green circles are drawn based on the center of the detected particles and a fixed radius. In panels (c, d), the wall particles are masked as black regions.}
\label{fig:exp}
\end{figure}

To relate macroscopic rheological behavior to microscopic particle motion, we conducted shear experiments on a quasi-2D granular raft, a monolayer of floating particles on a liquid surface (Fig.~\ref{fig:exp}(a), \cite{sasaki_origin_2025} for details).
Spherical hydrogel particles were floated on a 10~mm-deep transparent aqueous solution of sodium polytungstate (density $\rho$~= 2.6--2.9$\times 10^3$~kg\,m$^{-3}$, viscosity $\eta_\mathrm{f}$~= $2\times10^{-2}$~Pa\,s).
In the solution, all the hydrogel particles remained on the liquid surface, with a density of $\rho_\mathrm{p}$~= 2.6$\times 10^3$~kg\,m$^{-3}$ (slightly lower than $\rho$) and a mean diameter 4.3--4.4$\times 10^{-3}$~m (standard deviation 0.1$\times 10^{-3}$~m); hereafter, we use $d$~= 4.3$\times 10^{-3}$~m.
The low elastic modulus of the hydrogel \cite{gong_friction_2006} helped minimize out-of-plane motion of the spherical particles during the experiments.
To reduce surface tension and enhance the dispersibility, the solution density was adjusted slightly above that of the hydrogel spheres, and, except for the experiments with the widest channel system, $\sim$~1 wt\% of a surfactant (sodium alkyl ether sulfate) was added.

The experiments were carried out using a transparent Couette cell.
Shear was applied to the annular gap by rotating a motor at an angular rate of $\Omega$, which was connected to the inner wall via a torsion spring. 
The same particles were densely glued to both the inner and outer walls at the same height, with their centers of mass positioned at the radial distance from the container center $r=R_\mathrm{in}$ and $R_\mathrm{out}$, respectively.
In the widest channel system, particles were glued only to the inner wall, yet no motion was visible at $r=R_\mathrm{out}$ \cite{sasaki_origin_2025}.
Hereafter, the radial positions of the boundary walls are defined as the center of mass of each fixed particle, as shown in Fig.~\ref{fig:exp}(a).

Shear experiments were performed under various conditions of rotation rate $\Omega$~= $1\times10^{-3}$--1.3~rad\,s$^{-1}$, system size $R_\mathrm{in}$~= 2.7$d$--17$d$, channel width $w=R_\mathrm{out}-R_\mathrm{in}$~= 8.3$d$--32$d$, and packing fraction $0.67\le\phi\le0.81$ (Table~\ref{tb:run}). 
All shear experiments were conducted within the elastic regime of the hydrogel spheres.
During each experiment, the torque $\mathit{\Gamma}(r=R_\mathrm{in}-d/2)$ was measured at 2~Hz for each time $t$, using a B-type viscometer (BROOKFIELD, LVDV-II+Pro).
Simultaneously, the positions of all particles $\{\mathbf{r}_i(t)\}$ were recorded from the bottom of the container (Fig.~\ref{fig:exp} (b)), where $i$ labels the particles. Recording was performed at 1 or 10~fps for $R_\mathrm{in}\Omega$ below or above 1$\times 10^{-3}$~m\,s$^{-1}$, respectively, with a USB vision camera (OMRON SENTECH, STC-MCCM401U3V) and a machine vision lens (RICOH OPTOWL, FL-BC1220-9M).
To enhance the optical contrast, two panel lights were placed on each side of the system.
The effective resolution of the recorded images was $2048\times2048$ pixels, and a single pixel corresponds to 1.6$\times 10^{-4}$~m at the center ($d$ corresponds to approximately 27 pixels).
The distortion between the center and edge of the field of view was 2\% at maximum, while we did not apply any correction for optical refraction, distortion, and misalignment.
Using the measured torque and local velocity field, we investigated the representative flow law on the wall and the local flow law of the internal particles.

\begin{table}
\caption{Experimental and analysis conditions for the data used in the local velocity field analysis. \label{tb:run}}
\begin{ruledtabular}
\begin{tabular}{rrrrcl}

Run\#
&
$w/d$
&
$R_\mathrm{in}/d$
&
Particles\footnotemark
&
$\phi$\footnotemark
&
$3\Omega\Delta t/2\pi$\footnotemark
\\
\hline
97\hspace{4pt}  & 8.3\hspace{-7pt}  & 2.7& 273  & 0.69 & \hspace{10pt}1\\
99\hspace{4pt}  & 10 & 17\hspace{7pt} & 1074 & 0.67& \hspace{10pt}1\\
106\hspace{4pt} & 14 & 2.7  & 698  & 0.67& \hspace{10pt}1\\
108\hspace{4pt} & 17 & 17\hspace{7pt} & 2236 & 0.67& \hspace{10pt}1\\
102\hspace{4pt} & 25 & 2.7& 1897 & 0.67& \hspace{10pt}1, 15\\
82\hspace{4pt}  & 32 & 2.7 & 3147 & 0.69& \hspace{10pt}1, 15\\
80\hspace{4pt}  & 32 & 2.7 & 3265 & 0.72& \hspace{10pt}1, 5\\
74\hspace{4pt}  & 32 & 2.7 & 3333 & 0.73& \hspace{10pt}1, 15\\
83\footnotemark\hspace{-1pt}  & 32 & 2.7 & 3473 & 0.76& \hspace{10pt}1, 25\\
73\footnotemark  & 32 & 2.7& 3693 & 0.81& \hspace{10pt}1, 15\\
\end{tabular}
\end{ruledtabular}
\footnotetext[1]{Number of particles excluding those on the walls. The particles on the outer wall in the latter five runs are not fixed.}
\footnotetext[2]{Packing fraction defined by $d$~= 4.3$\times 10^{-3}$~m and the number of particles and their area within $R_\mathrm{in}+d/2<r<R_\mathrm{out}-d/2$.}
\footnotetext[3]{Time window $\Delta t$ for the velocity analysis. Additional longer intervals were applied at $\Omega$~= 1$\times10^{-2}$~rad\,s$^{-1}$ (Runs \#102, 82, 80, 83, 73), 1$\times10^{-1}$~rad\,s$^{-1}$ (Run \#74), and 1~rad\,s$^{-1}$ (Run \#102).}
\footnotetext[4]{Some particles located out of the raft plane.}
\footnotetext[5]{Many particles unavoidably located out of the raft plane, and thus an apparent $\phi$ is formally obtained.}

\end{table}

Using the recorded images (Fig.~\ref{fig:exp} (b--d)), all particle positions $\{\mathbf{r}_i(t)\}$ were tracked frame by frame with OpenCV software \cite{bradski_opencv_2000}. 
Details of this image-tracking analysis are provided in Appendix~\ref{sec:app}.
To obtain the radial distribution of the particle velocity $\mathbf{v}_i(t)$,
we first differentiated the position $\mathbf{r}_i(t)$ over a sliding time window $\Delta t$~= $2\pi/3\Omega$, one-third of the rotation period, and with an additional longer $\Delta t$ for prolonged experiments indicated in Table~\ref{tb:run}.
Using multiple time windows prevents any aliasing and improves the velocity resolution, which is $3\times 10^{-8}$~m\,s$^{-1}$ at best in this study and depends on $\Delta t$, while the uncertainty of the particle detection could make it a few times larger.
The switching of time-window intervals was performed within a region of $\Delta t$-independent velocity, providing a consistency check against resolution-limit artifacts and ensuring a smooth transition in the velocity.
Meanwhile, the variance of velocity strongly depends on $\Delta t$; therefore, it was not further considered in this study.
Second, the velocity $\mathbf{v}_i$ was averaged over the tangential direction with a radial width of $d/5$ at $r=md/10$ with $m$~= 0, 1, 2..., defining the coarse-grained velocity field $\mathbf{v}(r,t)= (v_\theta, v_r)$.
As the sheared granular raft exhibited fluctuations in the velocity as well as the torque owing to stick-slip behavior,
we further took their temporal averages over durations of 1--90 rotations, with fewer rotations averaged at slower $\Omega$ due to limited experimental duration.
Temporal averages were calculated from more than one rotation or more than 1000~s after the start for $\langle \mathbf{v} \rangle(r)$ and from when the torque reached steady state for $\langle\mathit{\Gamma}\rangle$.

\section{Results}\label{sec:res}
To obtain a constitutive relationship between stress and strain rate, we investigated the velocity field connecting to shear strain rate $\dot{\gamma}$ as,
\begin{equation}
\langle\dot{\gamma}\rangle(r) = \frac{\partial\langle v_\theta\rangle}{\partial r} - \frac{\langle v_\theta\rangle}{r}~.
\label{eq:strainrate}
\end{equation}
Shear stress $\tau$ can be analytically calculated from the measured torque with an assumption of angular momentum conservation as,
\begin{equation}
\langle \tau\rangle(r) = \frac{\langle \mathit{\Gamma}\rangle}{2\pi d r^2}~,
\label{eq:stress}
\end{equation}
which is shown in Fig.~\ref{fig:radial} (a).
In the following, we show the velocity field at various rotation rates, system sizes, and packing fractions under the quasistatic regime (Sec.~\ref{sec:vel}).
Using these results, constitutive flow laws are obtained (Sec.~\ref{sec:flowlaw}).

\subsection{Velocity field}\label{sec:vel}
\begin{figure}
\vspace{-1cm}
\includegraphics{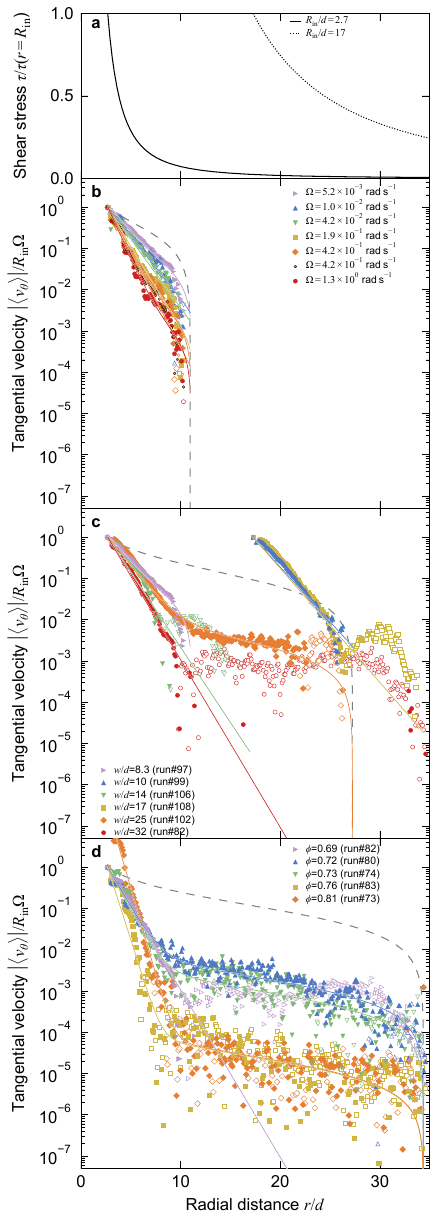}%
\caption{Radial distributions of flow. (a) Shear stress distributions in the systems with $R_\mathrm{in}/d$~= 2.7 (solid line) and 17 (dotted line), respectively. (b--d) Absolute tangential velocity $|\langle v_\theta\rangle|$.
Solid and open symbols represent the positive and negative mean velocities, respectively. 
Data were obtained under (b) various shear strain rates at $w/d$~= 8.3 and $\phi$~= 0.69 in Run \#97,
(c) various system sizes at $\dot{\gamma}(r=R_\mathrm{in})$~= 0.01--0.03 s$^{-1}$ and $\phi$~= 0.67--0.69, and (d) various packing fractions at $\Omega$~= $1.0\times10^{-2}$~rad\,s$^{-1}$ and $w/d$~= 32. The solid and dashed lines in panels (b--d) indicate the fittings with Eq.~(\ref{eq:exp}) and the analytical solutions for a Newtonian fluid with the system sizes of Runs \#97, \#102, \#83, respectively.
For reference, the same data of Run \#97 and \#82 are shown in panels (b--c) and (c--d), respectively.}
\label{fig:radial}
\end{figure}

The radial distribution of the particles' tangential velocity $\langle v_\theta \rangle(r)$ shows an exponential decay near the rotating cylinder, regardless of rotation rate $\Omega$ in Run \#97 (Fig.~\ref{fig:radial} (b)).
In this study, we call this innermost decay region a shear band, whose functional form is consistently observed in a bubble raft \cite{debregeas_deformation_2001} and a dense suspension \cite{huang_flow_2005}, as well as other various granular systems \cite{howell_fluctuations_1999,mueth_signatures_2000,komatsu_creep_2001,gdr_midi_dense_2004,tsai_granular_2005,fabich_measurements_2018}.
We have found that the following function can explain the velocity field, shown as solid lines in Fig.~\ref{fig:radial}(b), with a fitting parameter set ($\lambda$, $B$),
\begin{equation}
\langle v_\theta \rangle =\Omega R_\mathrm{in}\exp\left(-\frac{r-R_\mathrm{in}}{\lambda}\right)+BF(r)~.
\label{eq:exp}
\end{equation}
To be consistent with the boundary condition, $\mathbf{v}(r=R_\mathrm{out})=0$, the analytical solution of the velocity for the incompressible Newtonian fluids can be applied to $F(r) = C_1r+C_2/r$, where $C_1=-\Omega\tilde{R}^2/(1-\tilde{R}^2)$ and $C_2=\Omega R_\mathrm{in}^2/(1-\tilde{R}^2)$ with $\tilde{R}=R_\mathrm{in}/R_\mathrm{out}$ \cite{chandrasekhar_hydrodynamic_1961}.
$\langle v_\theta \rangle =F(r)$ for the Newtonian fluids is shown as a dashed line in Fig.~\ref{fig:radial} (b).
Note that $B$ is not necessarily nonzero in the best fits, shown as some simple exponential curves in Fig.~\ref{fig:radial} (b).
While the characteristic decay length $\lambda$ decreases and shear localizes with a increase in $\Omega$,
the velocity field exhibits a qualitatively similar pattern even at each time instant during torque fluctuation.

System size does not affect the emergence of shear bands with the exponential decay (Fig. \ref{fig:radial} (c)).
However, depending on the channel width $w$ under similar strain rate $\dot{\gamma}(\lambda)$ and packing fraction $\phi$, three types of radial zoning with a shear band are recognized.
At $w/d <11$ (Runs \#97, \#99), only the shear band occupies the channel.
At $11\lesssim w/d\lesssim20$ (Runs \#106, \#108), 
the outer region develops in which particles exhibited negative mean velocities and formed counter-flowing layers, shown as open symbols in Fig.~\ref{fig:radial} (c).
Alternatively, at $20\lesssim w/d$ (Run \#102), the outer particles create a creep region with slower decay than the shear band.
This creep region is recognized even at higher packing fractions (Fig.~\ref{fig:radial} (d)).
Note that Run \#82 is an exception owing to the loss of dispersibility caused by particle aggregation.
Equation (\ref{eq:exp}) with the analytical solution for the fluid well explains the creep region profile.

\begin{figure}
\includegraphics{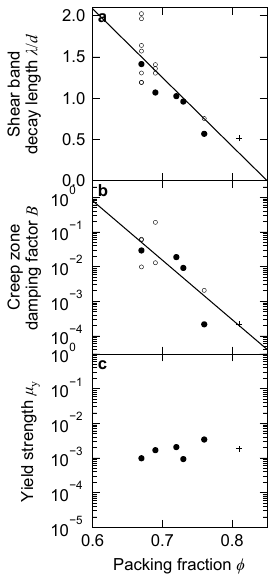}%
\caption{Dependence of (a) shear band decay length $\lambda$, (b) creep zone damping factor $B$, and (c) yield strength $\mu_\mathrm{y}$ on packing fraction $\phi$ in the quasistatic regime ($\langle I_\mathrm{in}\rangle<3\times 10^{-3}$). The solid symbols in panels (a, b) are shown at the same $\Omega$~= $1\times10^{-2}$~rad\,s$^{-1}$ and $R_\mathrm{in}/d$~= 2.7 for $w/d$~= 32 ($\phi>0.67$), with $w/d$~= 25 ($\phi$~= 0.67) shown for reference. The open symbols represent the other ($\Omega$, $R_\mathrm{in}$, $w$) conditions in the quasistatic regime. The solid lines represent the global fits to all data: $\lambda/d = -8.33\phi+7.08$ (a) and $B = \exp(-39.5(\phi-\phi_\mathrm{c}))$ with $\phi_\mathrm{c}$~= 0.594 (b). The data in panel (c) are mean values at each $\phi$, regardless of system size (Fig.~\ref{fig:muI}). Note that at $\phi$~= 0.81 (Run \#73) with many particles out of the raft plane, the cross symbols are used in panels (a--c).}
\label{fig:params}
\end{figure}

In contrast, packing fraction $\phi$ affects the velocity profile (Fig.~\ref{fig:radial} (d)) and its parameters in Eq.~(\ref{eq:exp}), $\lambda$ and $B$, 
in the quasistatic regime (discussed in Sec.~\ref{sec:flowlaw}).
Fig.~\ref{fig:params} (a), (b) show that variations in $\lambda$ and $B$ emerge depending on $\phi$ even at the same rotation rate $\Omega$~= $1.0\times10^{-2}$~rad\,s$^{-1}$ in the wider channels of $w/d$~= 32 or 25 (solid and cross symbols).
The shear band decay length $\lambda(\phi)$ linearly decreases from $1.5d$ to $0.5d$ as $\phi$ increases from 0.67 to 0.81.
Note that, for Run \#73, where the packing fraction was the highest and particles could not be confined to a single layer, $\langle v_\theta \rangle/R_\mathrm{in}\Omega>1$ near the inner wall surface, reflecting the formation of a plug undergoing rigid-body-like rotation with the inner wall owing to strong interparticle constraints (Fig.~\ref{fig:radial} (d)).

To account for this specific case only, we replaced $R_\mathrm{in}$ with $R_\mathrm{in}+1.5d$ in all subsequent calculations, including data fitting with Eq.~\ref{eq:exp}, where the factor 1.5 was determined empirically.
In any case, an increase in $\phi$ leads to shear localization (Fig.~\ref{fig:params} (a)).
Moreover, the creep zone damping factor $B$ exponentially decreases with $\phi$, enhancing the shear localization.
Using the local velocity distribution characterized by $\lambda$ and $B$, we can obtain the local flow laws as follows.

\subsection{Flow law}\label{sec:flowlaw}
Following the conventional formulation of flow law, we obtained a $\mu(I)$ rheology by scaling strain rate and stress in Eqs.~(\ref{eq:strainrate}) and (\ref{eq:stress}).
We calculated the apparent inertial number $\langle I\rangle(r)=-\langle \dot{\gamma} \rangle d/\sqrt{\langle{P}\rangle/\rho_\mathrm{p}}$ using the fitting function of $\langle v_\theta\rangle(r)$ (Eq.~(\ref{eq:exp})) in Eq.~(\ref{eq:strainrate}), and the apparent friction $\langle\mu\rangle(r)=\langle\tau\rangle/\langle P\rangle$ using Eq.~(\ref{eq:stress}).
A negative sign makes $I$ positive as $\langle \dot{\gamma}\rangle <0$ owing to the experimental geometry.
As measuring $P$ is difficult in a constant volume Couette cell in this study, we assumed hydrostatic pressure $\langle P\rangle=\rho_\mathrm{p}gd$ with gravitational acceleration $g$.
Note that $\langle P\rangle(r)$ is expected to be constant across the channel below the jamming transition point, which is also confirmed in a numerical simulation study \cite{koval_annular_2009}.

\begin{figure}
\includegraphics{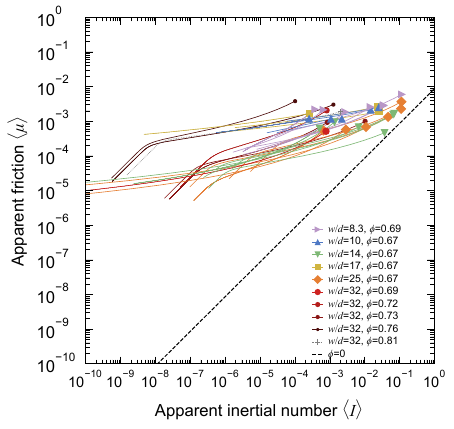}%
\caption{Relationship between the inertial number $\langle I\rangle$ (normalized shear strain rate) and friction (normalized shear stress) $\langle\mu \rangle$. Each symbol indicates the representative rheology measured on the inner wall ($r=R_\mathrm{in}$). The solid lines show the local $\langle\mu \rangle(\langle I\rangle(r))$ relationship obtained from the particle velocity field. The black dashed line indicates Newtonian reference for $\phi=0$.}
\label{fig:muI}
\end{figure}

Figure~\ref{fig:muI} shows the flow laws $\langle\mu\rangle(\langle I\rangle)$.
We found a discrepancy between them at $r=R_\mathrm{in}$ (symbols) and $r>R_\mathrm{in}$ (solid lines).
At $r=R_\mathrm{in}$, the relationship between $\langle \mu_\mathrm{in}\rangle=\langle \mu(r=R_\mathrm{in})\rangle$ and $\langle I_\mathrm{in}\rangle=\langle I(r=R_\mathrm{in})\rangle$ indicates a constant yield strength $\mu_\mathrm{y}$ in the quasistatic regime, consistent with the Herschel-Bulkley model \cite{bird_rheology_1983}, $\langle \mu_\mathrm{in}\rangle= \mu_\mathrm{y}+k'\langle I_\mathrm{in}\rangle^n$ with $n=1.3$ \cite{sasaki_shear_2025} and a flow consistency index $k'$.
Despite some scatter in $\langle \mu_\mathrm{in} \rangle$ depending on the system sizes, the mean value was obtained as $\mu_\mathrm{y}$ for each packing fraction $\phi$ (Fig.~\ref{fig:params} (c)) in the quasistatic regime, which is defined as $\langle I_\mathrm{in}\rangle<3\times10^{-3}$ hereafter and where no significant dependence of $\langle \mu \rangle$ on $\langle I \rangle$ was observed (Fig.~\ref{fig:muI}).
A decrease in $\phi$ slightly decreases $\mu_\mathrm{y}$ (Fig.~\ref{fig:params} (c)) and asymptotically reduces to a purely Newtonian behavior (at $\phi$~= 0) as shown by the dashed line in Fig.~\ref{fig:muI}.
In the quasistatic regime of $\mu=\mu_\mathrm{y}$ investigated here (Figs.~\ref{fig:radial} (c--d), \ref{fig:params}), our experimental system is characterized by a small Stokes number of $St\lesssim$ 0.01 with a flow time scale $\sim w/\Omega R_\mathrm{in}$, as well as a particle Reynolds number $Re_\mathrm{p}\simeq$ 0.1 and a Rayleigh layer depth $\delta\sim$ 0.01~m with a particle motion duration $\sim1~s$ \cite{sasaki_origin_2025}. 
This also leads to the expectation of quasistatic ($\langle\mu \rangle \ge\mu_\mathrm{y}$) or stationary ($\langle \mu \rangle=\mu_\mathrm{y}$) states of the particles.

At $r>R_\mathrm{in}$, however, the local flow law $\langle \mu\rangle(\langle I\rangle)$ exhibits a range of behaviors that deviate from $\langle \mu_\mathrm{in}\rangle(\langle I_\mathrm{in}\rangle)$, as shown by the multiple thin solid lines in Fig.~\ref{fig:muI}.
This variation arises from the parameters $\lambda$ and $B$ in Eq.~(\ref{eq:exp}), both of which are sensitive to the experimental conditions as discussed below (Fig.~\ref{fig:params}).
Notably, in the quasistatic regime, i.e., $\langle \mu_\mathrm{in}\rangle= \mu_\mathrm{y}$, the granular raft system exhibits nonzero velocities even at $\langle\mu\rangle<\mu_\mathrm{y}$, though only boundary slip at $r=R_\mathrm{in}$ is expected.
Therefore, the constitutive flow law for each local position exhibits spatial dependence, 
requiring a nonlocal effect that incorporates the influence of the surrounding flow.

\section{Discussion}\label{sec:disc}
The flow dynamics of granular rafts can be explained as a combination of diffusive granular motion in the shear band and competition between background shear and capillarity in the creep zone, as follows.
We established a universal flow law for the nonunique $\mu(I)$ relationships within the localized shear band under the quasistatic regime (Sec.~\ref{nonlocal}).
Furthermore, the dynamics for the outer creep region is discussed based on the heterogeneous instantaneous velocity field (Sec.~\ref{sec:creep}).

\subsection{Shear band's flow law: Rescaled $\mu(I)$ rheology} \label{nonlocal}
\begin{figure}[!tbp]
\includegraphics{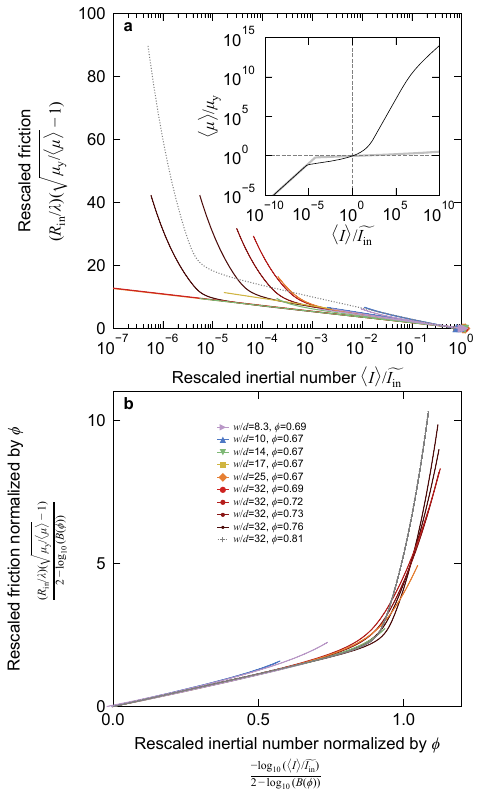}%
\caption{Universal flow law for the hydrogel granular raft in the quasistatic regime. (a) Relationship between the rescaled inertial number (strain rate) and the rescaled friction (stress). Data points near (1, 0) correspond to $r=R_\mathrm{in}$ or to flows without nonlocal effects, while all lines represent local flows with nonlocal effects. Note that smaller values on the y-axis correspond to larger values of $\langle \mu \rangle(r)$. Symbols and colors are as in panel (b).
The inset shows an overview of the constitutive flow law in Eq.~(\ref{eq:constitutive}). The solid and gray lines represent Run \#83 ($\Omega$~= 1$\times 10^{-2}$~rad\,s$^{-1}$) and the case with its $\lambda$ reduced to 10\%, respectively. The dashed lines indicate $\langle I \rangle/\widetilde{I_\mathrm{in}}=1$ and $\langle \mu \rangle/\mu_\mathrm{y}=1$ for comparison with Fig.~\ref{fig:muI}.
(b) Relationship in panel (a) normalized by $B(\phi)$.
Note that the parameters $\lambda$, $B$, and $\mu_\mathrm{y}$ use the fitted values for each condition (Fig.~\ref{fig:radial}) in panels (a), (b), and the inset of (a), except that $\lambda$ in (a) uses the global fitting function of $\phi$ from Fig.~\ref{fig:params} (a). The dotted line represents $\phi$~= 0.81 (Run \#73) with many particles out of the raft plane.}
\label{fig:univ}
\end{figure}

Within the shear band, we can obtain a constitutive law for the nonunique, local flow of granular rafts, starting with a universal fitting function in Eq.~(\ref{eq:exp}), though $\mu(I)$ cannot provide a unified framework to explain local flow laws in Fig.~\ref{fig:muI}.
As $v_\theta$ and $r$ in Eq.~(\ref{eq:exp}) can be converted to $I$ and $\mu$ through Eqs.~(\ref{eq:strainrate}), (\ref{eq:stress}), respectively,
we can rewrite Eq.~(\ref{eq:exp}) as,
\begin{widetext}
\begin{equation}
\langle I\rangle=\frac{d\sqrt{\rho_\mathrm{p}}}{\sqrt{P}} 
\left\{\Omega\left(\frac{R_\mathrm{in}}{\lambda}+\sqrt{\frac{\langle \mu \rangle}{\mu_\mathrm{y}}}\right)
\exp\left[-\frac{R_\mathrm{in}}{\lambda}\left(\sqrt{\frac{\mu_\mathrm{y}}{\langle \mu \rangle}}-1\right)\right]
+\frac{2BC_2}{R_\mathrm{in}^2}\frac{\langle \mu \rangle}{\mu_\mathrm{y}} \right\}~.
\label{eq:constitutive}
\end{equation}
\end{widetext}
Here, in general, we can also replace $\mu_\mathrm{y}$ with $\langle \mu_\mathrm{in} \rangle$.
At $r=R_\mathrm{in}$, $\langle\mu\rangle=\mu_\mathrm{y}$ and $\langle I_\mathrm{in}\rangle \simeq \widetilde{I_\mathrm{in}} \equiv d\sqrt{\rho_\mathrm{p}/P}\,\Omega(R_\mathrm{in}/\lambda+1)$ with $B \ll1$.
The functional form of Eq.~(\ref{eq:constitutive}) is shown as the solid black line in the inset of Fig.~\ref{fig:univ} using the parameters of Run \#83 at $\Omega$~= 1$\times 10^{-2}$~rad\,s$^{-1}$.
The local $\langle \mu\rangle(\langle I \rangle)$ relationship in Fig.~\ref{fig:muI} corresponds to the region of $\langle\mu\rangle/\mu_\mathrm{y}\le1$ and $\langle I \rangle/\widetilde{I_\mathrm{in}} \le 1$ in the inset of Fig.~\ref{fig:univ}.
When $\lambda$ is reduced to 10\% of the original value, the gray line shows the relationship with a constant yield strength $\mu_\mathrm{y}$ as with the global Herschel-Bulkley model except for the creep region.
Using the term in the exponent of Eq.~(\ref{eq:constitutive}), the various $\langle \mu\rangle(\langle I \rangle)$ relationships in the shear band can be collapsed in the framework of the relationship between the rescaled inertial number $\langle I \rangle/\widetilde{I_\mathrm{in}}$ and the rescaled friction $(R_\mathrm{in}/\lambda) (\sqrt{\mu_\mathrm{y}/\langle \mu\rangle}-1)$, except for the highest-$\phi$ case with many particles out of the raft plane (Run \#73), as shown in Fig.~\ref{fig:univ} (a).
Note that the fitting function of $\lambda(\phi)$ in the scaling factor is consistently used for each $\phi$ (solid line in Fig.~\ref{fig:params} (a)), while $\mu_\mathrm{y}$ is used as the original values owing to the scatter (data points in Fig.~\ref{fig:params} (c)).
This scaling consistently explains the locality-dependent flow law in the quasistatic shear band, which is independent of the assumption of $\langle P\rangle$.

To obtain the locality-dependent (i.e., nonlocal) flow law of Eq.~(\ref{eq:constitutive}) originated from Eq.~(\ref{eq:exp}),
the following heuristic equation, based on the concept originally proposed for emulsion systems \cite{goyon_spatial_2008}, has been found applicable in dry, dense granular systems \cite{bouzid_nonlocal_2013, bouzid_non-local_2015},
\begin{equation}
\langle I\rangle(r)=\langle I_\mathrm{in} \rangle(\langle\mu_\mathrm{in} \rangle=\langle \mu\rangle(r))+\xi^2 \nabla^2 \langle I\rangle(r)~.
\label{eq:diffusion}
\end{equation}
Here, we assume that $\mu_\mathrm{in}(I_\mathrm{in})$ corresponds to the representative $\mu(I)$ rheology, and $\xi$~= 0 means that the same flow law $\mu(I)$ holds throughout the entire system.
In the quasistatic regime, $\langle\mu_\mathrm{in} \rangle=\mu_\mathrm{y}$, leading to $\langle I_\mathrm{in} \rangle=0$ at $r>R_\mathrm{in}$.
This equation admits a general solution of exponential form, consistent with the first terms in Eqs.~(\ref{eq:exp}) and (\ref{eq:constitutive}); hence, $\xi \rightarrow \lambda$ at sufficiently large $r$ and small $\mu$.
Thus, a measure of nonlocal effects, $\lambda/R_\mathrm{in}$ in Eq.~(\ref{eq:constitutive}), can be regarded as the diffusion length scale of local granular motion $I$ \cite{bouzid_nonlocal_2013, bouzid_non-local_2015}, and can explain the shear band formation well below the yield strength \cite{wang_connecting_2020}.
This diffusion can be interpreted as the transfer of granular activity or fluidity, 
\cite{goyon_spatial_2008,henann_predictive_2013,bouzid_non-local_2015,kamrin_non-locality_2019}, which is defined as $I$ in this study.
The diffusion of $I$ may be driven by the roughened inner wall as a source of energy input in our system and by the resultant local jamming.
While the definition of fluidity and the physical meaning of Eq.~(\ref{eq:diffusion}) remain controversial \cite{bouzid_non-local_2015}, the concept of granular activity diffusion is supported by our observations of particle relative motion, dynamics slower than Newtonian flow, and the presence of yield strength, even in the quasistatic regime.

The diffusion length scale $\lambda/R_\mathrm{in}$ in Eq.~(\ref{eq:constitutive}) governs the shear band behavior.
It is controlled by strain rate and packing fraction (Fig.~\ref{fig:radial} (c), (d), respectively).
Increases in strain rate and packing fraction (Fig.~\ref{fig:params}) reduce $\lambda$ and enhance the shear localization, while a reduction of packing fraction, as discussed later in Sec.~\ref{sec:creep}, should result in Newtonian behavior.
Even under a small $\lambda$, however, its exponentially decaying flow, in principle, extends arbitrarily far \cite{komatsu_creep_2001} while the visible shear band of width ($\sim\lambda$) appears localized.
The limit $\lambda \to 0$ eventually results in a boundary slip, corresponding to the representative $\mu(I)$ flow law in the quasistatic regime without nonlocal effects.
A numerical simulation shows that an increase in packing fraction reduces the timescale of particle fluctuation, $d/\sqrt{\langle v_{\theta}^2 \rangle-\langle v_{\theta}\rangle^2}$, compared to that of bulk shear, $\mu/\dot{\gamma}$ \cite{zhang_microscopic_2017}, consistent with the observed reduction in the diffusion length scale $\lambda(\phi)$ (Fig.~\ref{fig:params} (a)).
In suspension rheology, instead of the diffusion of $I$, the square root of the viscous number $J=\eta_\mathrm{f}\dot{\gamma}/P$ \cite{cassar_submarine_2005,boyer_unifying_2011} or their sum $K=J+\alpha_\mu I^2$ with $1/\alpha_\mu$ being the transition Stokes number \cite{tapia_viscous_2022,tapia_rheology_2024,guazzelli_rheology_2024}
is also applicable to the shear-band (exponential) term in Eq.~(\ref{eq:constitutive}).

Therefore, our findings demonstrate that this nonlocality-incorporated rheology can apply not only to dry granular systems above the jamming transition point \cite{koval_annular_2009,kamrin_nonlocal_2012,kim_power-law_2020,gaume_microscopic_2020}
but also to dense suspensions below the jamming transition point \cite{lalieu_rheology_2023,huang_flow_2005} in the shear jammed state \cite{bi_jamming_2011}.

\subsection{Outer creep region's flow mechanism} \label{sec:creep}
\begin{figure}[!tbp]
\includegraphics{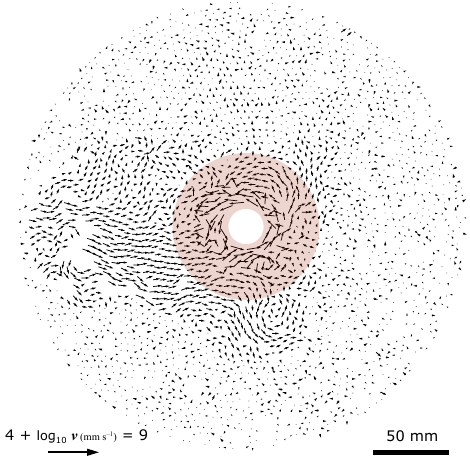}%
\caption{Instantaneous velocity field in Run \#80 (frame no. 1020). The plotted vector length is defined as $4+\log_{10}|\mathbf{v}|$, where $|\mathbf{v}|$ is in mm\,s$^{-1}$, to keep the vector direction and represent high and low velocities. Note that the velocity is calculated over time windows of $\Delta t$~= 200 and 1000~s for the regions $r \le 30$~mm and $> 30$~mm, respectively. The red hatched region corresponds to the shear band, where the velocity decays exponentially in the radial direction.}
\label{fig:velfield}
\end{figure}

The outer creep region can be phenomenologically accounted for by the analytical solution $F(r)$ for Newtonian flow with a damping factor $B$, while a unified framework for the shear band with Eq.~(\ref{eq:constitutive}) or (\ref{eq:diffusion}) is not applicable (Fig.~\ref{fig:univ} (a)).
The creep region is observed in wider channels even at high $\phi$ (Run \#83) and also in another granular raft in the quasistatic regime \cite{lalieu_rheology_2023};
hence this behavior might be common in extended granular rafts on the interface below the jamming point and justifies the use of the function $F(r)$ controlled by the interstitial and basal fluid flow.
However, the fluid flow $F(r)$ is damped by a factor $B$.
This velocity reduction may result from the effective increase in the Stokes number $St$, which is governed by the competition between the small background stress due to shear ($\propto r^{-2}$) and the inertial stress of particles aggregated by surface tension ($\sim$ constant).
The aggregation of particles can be confirmed as translational and rigid-body rotating clusters in the instantaneous velocity field in Fig.~\ref{fig:velfield} and as a deviation from the fit $F(r)$ in Fig.~\ref{fig:radial} (c), (d).
Microscopic dynamics of the clusters could be understood through further analysis of particle arrangement and velocity correlations, which are effectively captured by the factor $B$ in the present study.

The damping factor $B$ in Eq.~(\ref{eq:constitutive}) controls the creep zone behavior.
It depends on packing fraction (Fig.~\ref{fig:params} (b)) and little on strain rate.
An increase in packing fraction reduces $B$ probably owing to a decrease in interparticle distance and an increase in the size of the aggregated cluster, eventually leading to a channel-wide shear band described by the exponential flow law with fast decay.
In contrast, when packing fraction is sufficiently low, suspensions should exhibit Newtonian flow, following the Einstein or Krieger-Dougherty equation \cite{larson_structure_1999}.
As a decrease in $\phi$ increases not only $\lambda$ polynomially but also $B$ exponentially and even decreases $\mu_\mathrm{y}$ (Fig.~\ref{fig:params},
the first term in Eq.~(\ref{eq:diffusion}) from the local background contribution becomes significantly dominant over the second nonlocal term.
In this sense, the first and second terms in Eq.~(\ref{eq:diffusion}) correspond to the second linear (Newtonian) and first exponential terms in Eq.~(\ref{eq:constitutive}), respectively.

Furthermore, we found that the relationship between the logarithm of the inertial number and friction rescaled by $\lambda$ and $\mu_\mathrm{y}$ in Fig.~\ref{fig:univ} (a) can be globally scaled by the shifted order of magnitude of the damping factor, $2-\log_{10}{B(\phi)}$, as shown in Fig.~\ref{fig:univ} (b).
We confirmed that even a shift by one order of magnitude in the scaling factor would break the collapse.
Note that using each $B$ value (symbols in Fig.~\ref{fig:params} (b)) works well for the scaling in Fig.~\ref{fig:univ} (b), whereas the fitting function $B(\phi)$ (solid line in Fig.~\ref{fig:params} (b)) does not sufficiently collapse the data; more accurate measurements of $B(\phi)$ under well-dispersed conditions will be required.
Even with this limitation, as $B$ represents the deviation of $\phi$ from the critical value $\phi_\mathrm{c}\simeq$ 0.6 for $B$~= 1 and its two-order-of-magnitude variation corresponds to the difference of $\Delta \phi \sim 0.1$ (solid line in Fig.~\ref{fig:params} (c)),
the scaling factor $2-\log_{10}{B(\phi)}$ means the deviation from another critical value $\phi'_\mathrm{c}$~= 0.5.
The marginal regime of $\phi'_\mathrm{c}<\phi<\phi_\mathrm{c}$ might imply the competition between the shear localization involving nonlocal effects with large $\lambda$ and the Newtonian flow with $B$~= 1 and small $\mu_\mathrm{y}$, both of which govern the bulk flow behavior.

Even though the constitutive flow law in Eq.~(\ref{eq:constitutive}) continuously connects granular media and suspensions with the nonlocal effect to continuum viscous fluids without it,
the influences of fluid viscosity $\eta_\mathrm{f}$, distribution of packing fraction $\phi(I(r))$ and pressure $P(r)$, capillary force and resultant aggregation, and system geometry \cite{fenistein_wide_2003, fenistein_universal_2004,jagla_finite_2008,jop_hydrodynamic_2008, cromer_study_2014,  shukla_shear-banding_2019} warrant further investigations to test our hypothesis above.
In particular, higher-resolution measurements are needed to characterize both the viscous flow field of the interstitial and basal fluid and the inhomogeneity of the packing fraction within the shear band and the creep region.

While the detailed physical mechanisms remain to be clarified, our results imply that the contribution of the surrounding fluid to the overall system flow varies with the effective cluster size under aggregation.
In summary, the granular nonlocal effect manifests only at lower strain rates and intermediate packing fractions in the shear band, and the fluid flow becomes dominant at low packing fractions in suspension rheology.

\section{Conclusions}\label{sec:concl} 
In this study, we have experimentally investigated the flow law of a hydrogel granular raft suspension in the quasistatic regime.
We have found the universal local flow law for the shear band by incorporating the nonlocal size effect of the particle activity diffusion into $\mu(I)$ rheology.
The suspension flow in the shear band with nonlocal effects is well described and predicted by this constitutive flow law between the rescaled inertial number and the rescaled friction (Eq.~(\ref{eq:constitutive})) with the diffusion length scale $\lambda(\phi,\dot{\gamma})$, the viscous flow damping factor $B(\phi)$, and the yield strength $\mu_\mathrm{y}(\phi)$ (Fig.~\ref{fig:params}).
Thus, the established flow law is applicable to both dry and saturated granular flow above and below the jamming transition.
Outside the shear band, the creep region can be modeled as damped Newtonian fluid flow, which is inherent in multiphase suspension systems.
This study paved the way for establishing the universal constitutive flow law connecting the ductile/fluid flow to the brittle/granular behavior, which leads to developing the effective fluidization technology and assessing the granular fault activity in nature.

\appendix
\section{Method for particle tracking analysis} \label{sec:app}
We tracked all particles' positions by analyzing the recorded images.
First, the wall particle regions of thickness $d$ were masked.
We then segmented each particle region, by converting the original 8-bit images to grayscale, unsharp masking with Gaussian-blurred images (kernel size $51\times51$ pixels, $\sigma$~= 10) weighted by 0.6, adaptive thresholding using a Gaussian-weighted method (block size $501\times501$), median filtering (kernel size $3\times3$), and watershed segmentation (foreground obtained by thresholding the distance transform at 40\% of its maximum; background identified by 1-iteration dilation using a $3\times3$ kernel).
In each segmented binary image (Fig.~\ref{fig:exp} (c)), the center of each circular particle was detected with labels at the original image resolution, by Hough Circle Transform (minimum distance between detected circle centers 20 pixels, Canny edge detection threshold 100, accumulator decision threshold 4, circle radii between 11 and 13 pixels), as shown in Fig.~\ref{fig:exp} (d).
We performed particle tracking by associating particle positions and labels between consecutive frames using nearest neighbor search with a distance threshold of 10 pixels per frame.

\begin{acknowledgments}
Y. S. discloses support for the research of this work from the Association for Disaster Prevention Research [the Early-Career Scientist Research Grant 2024-5]. H. K. discloses support for the research of this work from JSPS [KAKENHI Grant Number JP24H00196] and JST [ERATO Grant Number JPMJER2401].
\end{acknowledgments}

\bibliography{references.bib}

\end{document}